\documentstyle[epsf,prl,aps]{revtex}

\newcommand{\be}{\begin{equation}}
\newcommand{\ee}{\end{equation}}

\newcommand{\Integer}{\mbox{$\bf \cal{Z}$}}
\newcommand{\ignorar}[1]{}

\title{Multifractal wavelet filter of natural images}

\author{Antonio Turiel\hspace*{4.5cm}N\'estor Parga\\
Laboratoire de Physique Statistique\hspace*{1.75cm}
Departamento de F\'{\i}sica Te\'orica\\
\hspace*{0.5cm}Ecole Normale Sup\'erieure\hspace*{1.5cm}
Universidad Aut\'onoma de Madrid\\
\hspace*{-1cm}24, rue Lhomond. 75231 Paris Cedex
05\hspace*{1cm}Cantoblanco 28049.
Madrid\hspace*{3cm}\\
France\hspace*{6cm}Spain\\
e-mail: Antonio.Turiel@lps.ens.fr\hspace*{1.5cm}e-mail:
parga@delta.ft.uam.es\hspace*{0.25cm}
}

\begin{document}

\twocolumn

\maketitle

\begin{abstract}
Natural images are characterized by the multiscaling properties of
their contrast gradient, in addition to their power spectrum. In this
work we show that those properties uniquely define an {\em intrinsic
wavelet} and present a suitable technique to obtain it from an
ensemble of images.  Once this wavelet is known, images can be
represented as expansions in the associated wavelet basis. The
resulting code has the remarkable properties that it separates
independent features at different resolution level, reducing the
redundancy, and remains essentially unchanged under changes in the
power spectrum. The possible generalization of this representation to
other systems is discussed.
\end{abstract}

Pacs: 87.57.Nk,47.53.+n,87.19.Dd,42.66.Lc

{\it Physical Review Letters}, {\bf 85}, 3325-3328 (2000)

\section{Introduction}

Given the complexity and degree of redundancy of natural images, the
early visual system had to find good coding procedures to represent
the visual stimuli internally.  To achieve this goal, the visual
system must have learnt the regularities present in the environment
where the organism lived \cite{Ba61}. If there are image features that
tend to appear together, a cell responding quasi-optimally to them is
rather likely to exist.  To find such a representation one has first
to understand the statistical properties of visual scenes common in
the environment.  In particular, the relevance of the second order
statistics has been pointed out some time ago \cite{Fi87}, and
internal representations that eliminate these correlations have been
discussed \cite{At92vHa92}.  However, even if whitening represents an
improvement of the code, it still leaves much geometrical structure
that should be dealt with more properly \cite{Fi93}.  A more
systematic study of statistical regularities that go beyond the
two-point correlations has began rather recently
\cite{RuBi94Ru94,TuMaPaNa98NIPS97,Singularities}.  A novel approach to
understand the statistical properties of natural images has been
proposed in \cite{TuMaPaNa98NIPS97} where the non-gaussian statistics
of changes in contrast has been characterized and explained by means
of a stochastic multiplicative process: Contrast changes at a given
scale are obtained from those at a coarser scale by multiplication
with an independent random variable. This implies a linear relation
between the logarithms of the variables at two different scales (this
property has also been discussed in \cite{BuSi99}, although the
existence of a multiplicative process was not noticed). 
A very rich geometrical structure has emerged from those 
studies: contrast changes are organized in such a way that pixels in
the image can be classified according to the strength of the
singularities of the contrast gradient \cite{Singularities}. It was
also checked that the multiplicative process is present in very
different sets of images and in color natural images 
\cite{NeTuPa99TuPaRuCr99TuPo99}.

In this work we show that when those findings are taken into account
in an image model there appears a very compact representation of the
visual world. In particular, a wavelet filter that guarantees that
variations in contrast at different scales are related by a
multiplicative process can be derived experimentally from a dataset of
natural images, with the only assumption of the existence of such a
filter. Furthermore, the code that it defines
has eliminated a great deal of redundancy.

\section{Wavelet basis and multiresolution analysis}

\indent 
We start by considering the projection of the contrast $c(\vec{x})$
on a dyadic wavelet set  $\tilde{\Psi}_{j\vec{k}}(\vec{x}) =
\tilde{\Psi}(2^j\vec{x}-\vec{k})$. The wavelet $\Psi_{j\vec{k}}$
focuses on the details of the image at the scale $r$, where $r=2^{-j}$,
$j\in\Integer$, at the sampling points $\vec{x}_0=2^{-j}\vec{k}$. 
Here $\vec{k}\equiv(k_1,k_2)$, with $k_1,k_2\in\Integer$\cite{Daubechies}.
If $\tilde{\Psi}$ defines a wavelet basis, the {\bf wavelet projections} 
$T_{\tilde{\Psi}}^r c(\vec{x}_0)$ of the field of luminosity contrast,

\be
T_{\tilde{\Psi}}^r c(\vec{x}_0)
\; =\;
r^{-2} \int d\vec{x} \;\tilde{\Psi}_r(\vec{x}-\vec{x}_0) c(\vec{x})
\; =\;
r^{-2} \langle\tilde{\Psi}_{r,\vec{x}_0}|c\rangle
\ee

\noindent
characterize completely the image. This is the reason
of the name ``multiresolution analysis'': the wavelet projection is a
description of $c(\vec{x})$ at the point $\vec{x}_0$ when the image is
observed at a variable scale $r$. This type of analysis reaches a
compromise between localization in position and in spatial
frequency \cite{Daubechies}. If the discrete basis is complete, then the
contrast can be expanded in a wavelet basis orthogonal to $\tilde{\Psi}$
using the wavelet projections as coefficients. The dual basis
$\Psi_{j\vec{k}}$ verifies that:

\be
\langle \tilde{\Psi}_{j\vec{k}}|\Psi_{j^{\prime}\vec{k}^{\prime}}\rangle
\; =\; 2^{-2j} \delta_{jj^{\prime}}\delta_{\vec{k}\vec{k}^{\prime}}
\label{eq:ortho}
\ee

\noindent
and the contrast $c(\vec{x})$ can then be expressed as:

\be
c(\vec{x})\; =\; \sum_{j,\vec{k}} 
\alpha_{j\vec{k}} \:\Psi_{j\vec{k}}(\vec{x})
\label{eq:wv_expansion_gral}
\ee

\noindent
where $\alpha_{j\vec{k}} = T_{\tilde{\Psi}}^{2^{-j}} c(2^{-j}\vec{k})
=2^{2j} \langle \tilde{\Psi}_{j\vec{k}}|c\rangle$. 
The wavelet basis generated by $\Psi$ will be called the
{\bf representation basis}. Its dual basis, defined by $\tilde{\Psi}$
will be referred to as the {\bf analizer basis}. Once one of these basis is
known, the other is completely defined by eq.~(\ref{eq:ortho}).

\subsection{Multiscaling in natural images}

In \cite{Singularities} it was experimentally proved that there exists
a certain class of wavelets, such that some of their low order moments
are zero, for which:

\be
\langle |T_{\tilde{\Psi}}^r c|^p \rangle
\; =\; \alpha_p\: r^{\tau_p^c}
\label{eq:SS_contrast}
\ee

\noindent
(the angular brackets denote the average over an ensemble of images),
a property known as Self-Similarity (SS). It was also found 
\cite{TuMaPaNa98NIPS97} that the SS exponents $\tau_p$ have a non-trivial
dependence on $p$ which can be explained by means of a stochastic
{\bf multiplicative process} (see e.g. \cite{Fr95}). In terms of
the wavelet coefficients $\alpha_{j\vec{k}}$, this process is
expressed as:

\be
\alpha_{j\vec{k}}
\; =\; \eta_{j\vec{k}}\:
\alpha_{j-1,\left[\frac{\vec{k}}{2}\right]}
\label{eq:reletaalpha}
\ee

\noindent
where $\left[\vec{\kappa}\right]$ denotes the vector with components
given by the integer part (rouding down) of those of $\vec{\kappa}$.
In \cite{TuMaPaNa98NIPS97,Singularities} eq.~(\ref{eq:reletaalpha}) was
understood in the statistical sense, where the variables
$\eta_{j\vec{k}}$ are statistically independent of the
$\alpha_{j-1,\left[\frac{\vec{k}}{2}\right]}$.  The SS exponents
$\tau_p$ define completely the distribution of the random variables
$\eta_{j\vec{k}}$ and vice-versa. Besides, the distribution of the
$\eta$'s depends only on the ratio between the two scales of the
wavelet projections \cite{TuMaPaNa98NIPS97}.  Here this ratio has been fixed
at 2 for any pair of consecutive scales.  This implies that all
the $\eta_{j\vec{k}}$'s are identically distributed.

\indent 
In this work we go beyond those results by showing that
natural images posses an intrinsic wavelet for which
eq.~(\ref{eq:reletaalpha}) is fulfilled {\it point by point}. This
means that the equality holds for any image, at any scale and position
and that the variables $\eta_{j\vec{k}}$ can be extracted directly
from:

\be
\eta_{j\vec{k}}\; =\;
\frac{\alpha_{j\vec{k}}}{\alpha_{j-1,\left[\frac{\vec{k}}{2}\right]}}
\label{eq:etaindep}
\ee

\indent 
In order to obtain these variables and to verify their
statistical properties, one has first to determine this wavelet. Under
the assumptions that the $\eta$'s obtained from
eq.~(\ref{eq:etaindep}) are scale independent and equally distributed
variables, the representation wavelet $\Psi$ can be experimentally
obtained from a statistical analysis of the image ensemble. This will
be shown in the next section. The validity of these two hypothesis on
the $\eta$'s has to be verified {\it a posteriori}, once the wavelet
is known. This is done in Section~\ref{section:project} as follows:
first the the analizer wavelet $\tilde{\Psi}$ is obtained from
the representation wavelet $\Psi$. In turn, $\tilde{\Psi}$ can be used
to evaluate the coefficients $\alpha_{j\vec{k}}$'s and from them the
$\eta_{j\vec{k}}$'s. Once these are known it is finally checked that
they are indeed scale-independent, identically distributed random
variables, so checking the self-consistency of the multiplicative
process model.

\begin{figure}[htb]
\begin{center}
\hspace*{0cm}
\leavevmode
\epsfxsize=5cm
\epsffile{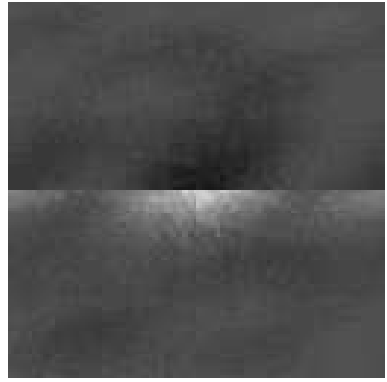}
\end{center}
\caption{ Representation wavelet $\protect\Psi$ for the image
ensemble. The function is represented in gray levels, the darkest points
indicate the sites where it takes its smallest values.  }
\label{fig:wave-vH}
\end{figure}

\section{The representation mother wavelet}
\label{section:mother_wv}

\indent

We suppose that the contrast field obtained from our image dataset can
be expanded as a superposition of wavelets, eq.(\ref{eq:wv_expansion_gral}),  
and that the $\alpha_{j\vec{k}}$'s verify eq.~(\ref{eq:reletaalpha}),
where the $\eta_{j\vec{k}}$'s are scale independent equally
distributed variables.  Since the images have finite size we will take
$j\geq 0$ where $\Psi_{0\vec{0}}$ covers the whole image and it
represents the mother wavelet, $\Psi_{0\vec{0}}\equiv\Psi$. For the
same reason, the range of $\vec{k}=(k_1,k_2)$ at the scale $j$ is
bounded as: $k_1,k_2=0,1,..., 2^j-1$.

\indent 
Averaging $c(\vec{x})$ at each point $\vec{x}$, it is found
that the average contrast can be represented as a simple wavelet
superposition: ${\cal C}(\vec{x})\equiv \langle c\rangle
(\vec{x})\propto \alpha_{0\vec{0}} \:\sum_{j,\vec{k}} {\overline{|\eta|}}^j
\Psi_{j\vec{k}}(\vec{x})$. Here $\overline{|\eta|}$ is the first order
moment of the distribution of the $|\eta|$'s, and we have used the
assumptions that all the $\eta_{j\vec{k}}$'s have the same marginal
distribution and are independent across the scales. By Fourier
transforming this field, $\hat{\cal C} (\vec{f})$, one easily obtains
the Fourier transform of the representation wavelet,
$\hat{\Psi}(\vec{f})$, that reads:

\be
\hat{\Psi}(\vec{f})\; =\; \frac{1}{\alpha_{0\vec{0}}} 
\left[ \hat{\cal C}(\vec{f})
-\frac{\overline{|\eta|}}{4}\;\frac{\Lambda(\vec{f})}
{\Lambda(\frac{\vec{f}}{2})} \hat{\cal C}(\frac{\vec{f}}{2})
\right]
\label{eq:mother_wv}
\ee

\begin{figure}[bht]
\begin{center}
\hbox{
	\hspace*{0cm}
        \leavevmode
        \epsfxsize=4cm
        \epsfbox[50 50 410 302]{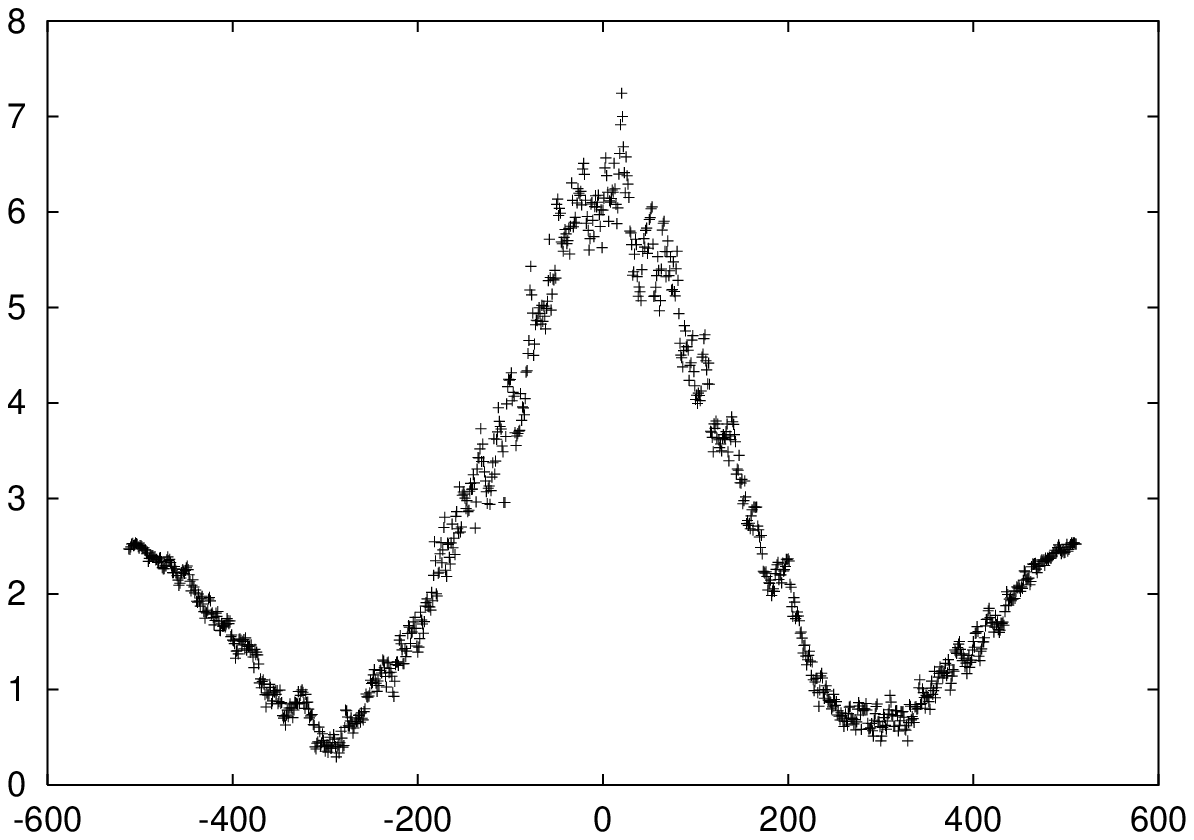}
        \epsfxsize=4cm
        \epsfbox[50 50 410 302]{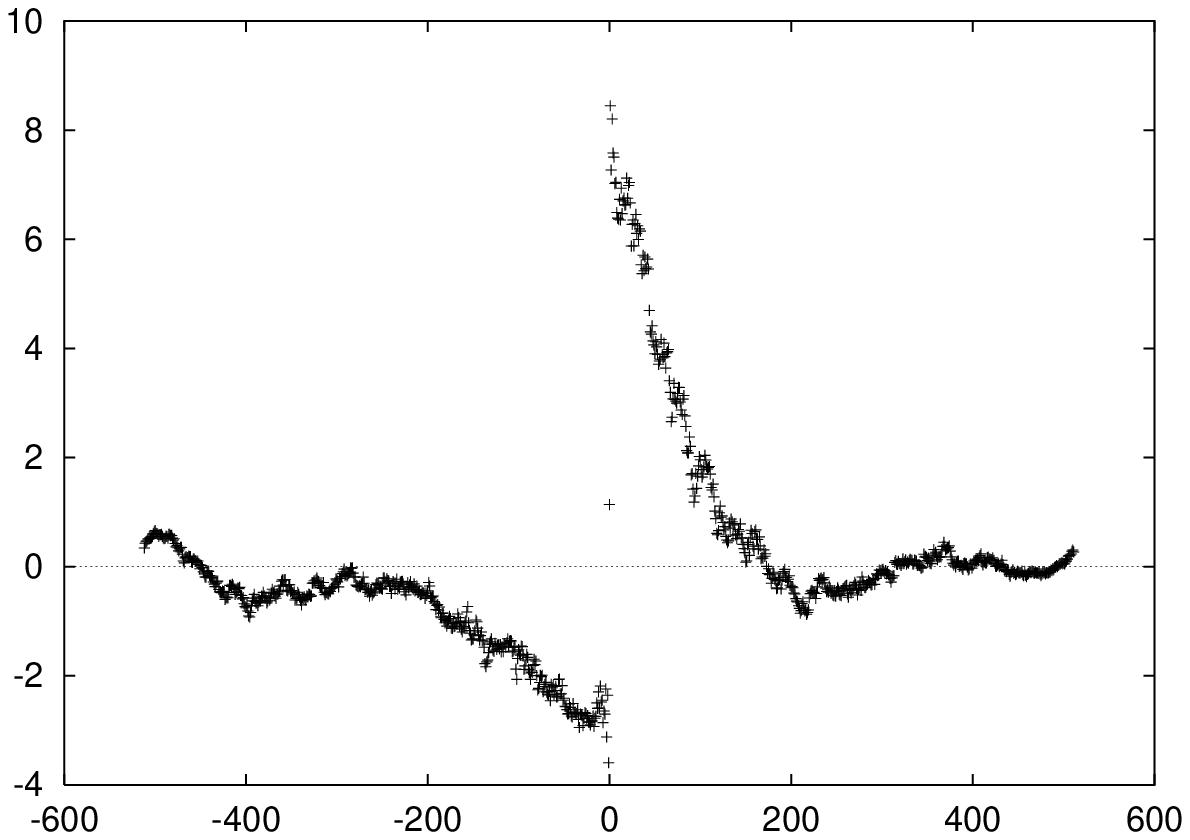}
}
\end{center}
\caption{The wavelet $\protect\Psi$ along the horizontal (left) and the
vertical axis (right).} 
\label{fig:wave-vH-1D}
\end{figure}

\noindent
where $\Lambda(\vec{f})=(1-e^{-2 \pi i f_1})(1-e^{-2 \pi i
f_2})$. This expression is very appealing. The right
hand side compares the average contrast at two consecutive scales
(related by a factor $2$), and it expresses that the wavelet is
obtained as an observation of the scale transformation properties of
the images. The average $\overline{|\eta|}$ has an {\it a priori} known value:
$\overline{|\eta|}=\frac{1}{2}$ \cite{Singularities}. This allows to
use eq.~(\ref{eq:mother_wv}) with no {\it a priori} knowledge about
the data.

\indent To obtain experimentally the representation wavelet,
eq.~(\ref{eq:mother_wv}) was applied to a large ensemble of natural
images. The data were 200 $1024\times 1024$ images selected at random
from van Hateren's dataset
\cite{Hateren_basis}. Figure~\ref{fig:wave-vH} shows the
representation wavelet obtained with this procedure. It exhibits two
clear features: it is right-left symmetric and roughly up-down
antisymmetric with a sharp central discontinuity (see
Figure~\ref{fig:wave-vH-1D}). The first property is expected: our
world remains statistically unchanged when it is reflected from left
to right. The rough up-down antisymmetry and the related discontinuity 
are probably due to the sharp contrast between the sky and the ground.

\section{The projection into the experimental basis}
\label{section:project}

\indent

Once the analizer wavelet is obtained, it can be used to evaluate the
coefficients $\alpha_{j\vec{k}}$ and from them, using
eq.~(\ref{eq:etaindep}), the coefficients $\eta_{j\vec{k}}$. Then, it
is possible to check whether the $\eta_{j\vec{k}}$ are
scale-independent, identically distributed variables. In the
affirmative case, this fact self-consistently demonstrates that
eq.~(\ref{eq:mother_wv}) is the intrinsic wavelet we are looking for.
It is immediate to check that the signs of the $\eta$'s are
independent of their absolute values, $|\eta|$, and also
scale-independent, so it is only necessary to verify the independence
of the $|\eta|$'s.

\indent 
The mutual dependence among the $|\eta_{j\vec{k}}|$'s was
estimated by computing the correlation coefficients between two
$|\eta|$'s. As the number of possible pairs is very large, only two
types of correlations were considered, which should be maximal by
construction: those between consecutive scales ($j-1$ and $j$) at
equivalent positions, and those between consecutive spatial positions
($\vec{k}$ and $\vec{k}+\vec{d}$, where the components of $\vec{d}$
take only the values $0$ and $1$) at a fixed scale $j$.

\begin{figure}[htb]
\begin{center}
\hspace*{0cm}
\epsfysize=3.5cm
\epsfbox[50 50 410 302]{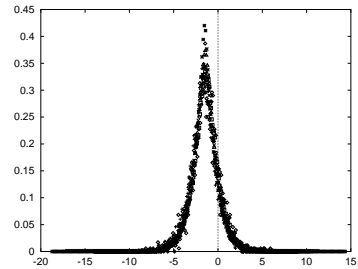}
\end{center}
\caption{P.d.f.'s of $\ln|\eta_{j}|$ at the scales $j=$
2 (diamonds), 3 (crosses), 4 (squares), 5 (x), 6 (triangles) and 7
(crossed x). }
\label{fig:etadist}
\end{figure}

\indent Both correlation coefficients, denoted as $\rho_j$ and
$\rho_{j\vec{d}}$ respectively, take values in $[-1,1]$ and give a
dimensionless measure of the degree of statistical dependence. The
observed values of $\rho_j$ are very close to $0$
($|\rho_j|<10^{-2},\: j>2$), what confirms rather well the
independence of the $\eta_{j\vec{k}}$'s under changes of scale. On the
contrary, the correlation coefficients $\rho_{j\vec{d}}$ are by no
means negligible, although they are extremely short-ranged: the
variables $\eta_{j\vec{k}}$ and $\eta_{j\vec{k}^{\prime}}$ are
independent when they are more than one pixel apart; after that
distance the two-point correlation $\rho_{j\vec{d}}$ decays
dramatically. It is important to remark that there is no need of
spatial independence of the $\eta_{j\vec{k}}$'s in our wavelet
model. For this reason, although short-ranged, the observed dependence
is a significant source of information about the remaining statistical
structure at a fixed resolution layer.  On the other hand, the
$\eta$'s define a system of almost completely independent variables.

\indent 
To confirm the validity of the wavelet representation,
eqs.~(\ref{eq:wv_expansion_gral}) and (\ref{eq:mother_wv}), 
one has still to check that the $\eta$'s are identically distributed.
Figure \ref{fig:etadist} exhibits the distribution of $\ln|\eta_{j}|$ for
different scales $j$. The correspondence between them is really very good.

\section{Wavelet transparency and decorrelation}

\indent

It has been frequently argued that the the signal that arrives to the
primary visual cortex has been already decorrelated in previous stages
of the visual system \cite{At92vHa92,DaAtCl96}.  In this case, V1
would take care of coding more complex aspects of images.  In this
regard, the way in which the mother wavelet is constructed guarantees
a remarkable property of transparency to the power spectrum: it
defines a code that is somewhat independent of the second order
statistics of the images.  The power spectrum of natural images
exhibits a power law behaviour, $S(\vec{f})\sim f^{-(2-\epsilon)}$
\cite{Fi87}. Under the assumption of translational invariance, the
application of the decorrelating filter to the contrast is equivalent
to multiplication in the Fourier domain by
$f^{1-\frac{\epsilon}{2}}$. Denoting the decorrelated contrast as
${\bf D}c(\vec{x})$, it is immediate to see that it has a
representation similar to eq.~(\ref{eq:wv_expansion_gral}):

\be
{\bf D}c(\vec{x}) \; =\; \sum_{j\vec{k}} \alpha_{j\vec{k}}^{\prime}\:
{\bf D}\Psi_{j\vec{k}}(\vec{x})   \;\;\;  ,
\ee

\noindent
where ${\bf D}\Psi$ indicates the application of the decorrelating
operator to the representation wavelet and
$\alpha_{j\vec{k}}^{\prime}= 2^{j(1-\frac{\epsilon}{2})}
\alpha_{j\vec{k}}$.  Defining now
$\eta_{j\vec{k}}^{\prime}=2^{1-\frac{\epsilon}{2}}\eta_{j\vec{k}}$ it
is concluded that the decorrelated images also posses 
a random multiplicative process. The new
representation wavelet ${\bf D}\Psi$ can be obtained from an
ensemble of decorrelated images by means of eq.~(\ref{eq:mother_wv}).
It is known that the value of the exponent $\epsilon$ has fluctuations
from image to image \cite{Tollhurst}.  The relevance of the wavelet
transparency is that if the visual stimulus has been decorrelated
before it reaches V1, this area can use the type of filters described
in this paper regardless of the precise exponent of the power
spectrum.

\section{Conclusions}

\indent 

We have shown that images can be represented as multiresolution
objects in terms of an appropriate wavelet basis $\Psi$, in which each
resolution level is an independent image.  One of the advantages of
this representation is that it is based on the observed properties of
the contrast gradient \cite{TuMaPaNa98NIPS97,Singularities}, what in
turn leads to an automatic reduction of the redundancy. At the same
time the spatial correlations at a given scale are short-ranged, but still
informative.

\indent 
Once the wavelet is known, it is possible to devise a compression
technique based on the properties of this representation.  This would
have two important advantages with respect to other wavelet
representations (see e.g. \cite{SaPe96}): first that the scale layers
are independent, and second that there exists a simple model for the
distribution of coeficcients
\cite{TuMaPaNa98NIPS97,Singularities,NeTuPa99TuPaRuCr99TuPo99} that
can be used in the coding.

\indent 
These results have been obtained with the simplest wavelet
expansion, but the tools presented in this work can be taken as the
starting point to look for more realistic visual filters.  This search
should be directed by the experimental observation that cells in V1
are edge detectors.  From this perspective, the expansion in
eq. (\ref{eq:wv_expansion_gral}) should be generalized to include an
orientational degree of freedom. Filters of this type have been
proposed in \cite{Da89} and found from an independent component
analysis of natural images \cite{BeSe97}
. These studies should be combined with the use of overcomplete basis,
this introduces redundancy but the representation becomes stable under
small changes in the images \cite{SiFrAdHe92OlFi97}

\indent 
This analysis could also be carried out over any physical
system with multiscaling properties, eq.~(\ref{eq:SS_contrast}) (for
instance, turbulent flows in fully developed turbulence \cite{Fr95}). For
all these systems eq.~(\ref{eq:mother_wv}) could be applied to obtain the
associated representation wavelet $\Psi$.  However, it would be necessary
to check the statements of scale independence and identical distribution
for the projection coefficients $\eta_{j\vec{k}}$. If both properties
hold, the compact code so obtained would be a valuable tool, and the
interpretation of the possible remaining structure at each scale very
meaningful.

\subsection*{Acknowledgements}
This work was funded by a Spanish grant PB96-0047. A. Turiel is
financially supported by a fellowship from the Spanish Ministry of
Education.


\begin{thebibliography}{99}

\bibitem{Ba61}
Barlow~H. B.,
in {\it Sensory Communication} (ed. Rosenblith W.)  pp. 217.
(M.I.T. Press, Cambridge MA, 1961)

\bibitem{Fi87} 
Field D. J., 
{\it J. Opt. Soc. Am.} {\bf 4}: 2379-2394 (1987)

\bibitem{At92vHa92} 
Atick~J. J.  
{\it Network} {\bf 3}: 213-251 (1992); 
van Hateren~J.H.
{\it J. Comp. Physiology A} {\bf 171}: 157-170 (1992)

\bibitem{Fi93}
Field D. J.,
in {\it Wavelets, Fractals, and Fourier Transforms}. 
Eds. Farge M., Hunt J.C.R. \& Vassilicos J.C., pp. 151-193.
Clarendon Press, Oxford (1993).

\bibitem{RuBi94Ru94}
Ruderman D. \& Bialek W.,
{\it Phys. Rev. Lett.} {\bf 73}: 814 (1994); 
Ruderman D.,
{\it Network} {\bf 5}: 517-548 (1994)

\bibitem{TuMaPaNa98NIPS97} 
Turiel A., Mato G., Parga N. \& Nadal J.-P. 
{\it Phys. Rev. Lett.} {\bf 80}: 1098-1101 (1998); 
Turiel A., Mato G., Parga N. \& Nadal J.-P.
{\it Proc. of NIPS'97}. MIT Press, Cambridge, MA (1998).  


\bibitem{Singularities} 
Turiel A. \& Parga N. 
{\it Neural Computation} {\bf 12}: 763-793 (2000)

\bibitem{NeTuPa99TuPaRuCr99TuPo99} 
Nevado A., Parga N \& Turiel A. 
{\it Network} {\bf 11}: 131-152 (2000); 
Turiel A., Parga N., Ruderman D. \& Cronin T.W.,
To appear in {\it Phys. Rev. E} (1999); 
Turiel A. \& del Pozo A. 
Sub. to IEEE Transactions on Image Processing (2000) 


\ignorar{
\bibitem{Novikov} 
Novikov, 
{\it Phys. Rev. E} {\bf 50}: R3303 (1994)
}


\bibitem{BuSi99} Buccigrossi R.W. and Simoncelli E.P.,
{\it IEEE Trans Image Processing} {\bf 8}, 1688-1701, (1999).


\bibitem{Daubechies} 
Daubechies I. 
{\it ``Ten Lectures on Wavelets''}.
CBMS-NSF Series in Ap. Math. {\it Capital City Press}. Montpelier, 
Vermont (1992)


\bibitem{Fr95}  
Frisch U., 
{\it Turbulence}, Cambridge Univ. Press (1995)


\ignorar{
\bibitem{Du94Ca96} 
Dubrulle B., 
{\it Phys. Rev. Lett.} {\bf 73}: 959-962 (1994); 
Castaing,  
{\it J. Physique II, France} {\bf 6}: 105-114 (1996)
}

\bibitem{Hateren_basis} 
Hateren, J.H. van \& Schaaf, A. van der
{\it Proc.R.Soc.Lond. B} {\bf 265}: 359-366 (1998)

\bibitem{DaAtCl96} 
Yang Dan, Atick J. \& Clay Reid R., 
{\it J. of Neuroscience} {\bf 16}: 3351-3362 (1996)

\bibitem{Tollhurst} 
Tollhurst D. J., Tadmor Y. \& Tang Chao,
{\it Ophthal. Physiol. Opt.} {\bf 12}: 229-232 (1992)

\bibitem{SaPe96}
Said A  \& Pearlman W
{\it IEEE Trans Image Proc} {\bf 5} 1303-1310 (1996)


\bibitem{Da89} 
Daugman J.G., ,
{\it IEEE Transactions on Biomedical Engineering} {\bf 36}: 107-114 (1989)

\bibitem{BeSe97} 
Bell A.J. \& Sejnowski T.J.,
{\it Vision Research} {\bf 37}: 3327-3338 (1997)

\bibitem{SiFrAdHe92OlFi97} 
Simoncelli E.P., Freeman W.T., Adelson E.H. \& Heeger D.J.,
{\it IEEE Transactions on Information Theory} {\bf 382}: 587-607 (1992); 
Olshausen B.A. \& Field D.J.,
{\it Vision Research} {\bf 37}: 3311-3325 (1997)




\ignorar{


\bibitem{Mallat} Mallat S., Zhong S. 
In {\it ``Wavelets and their applications''},
{\it Jones and Bartlett Publishers}. Boston 1991.



\bibitem{barlow61} Barlow~H. B., 
in {\it Sensory Communication} (ed. Rosenblith W.)  pp. 217.
 (M.I.T. Press, Cambridge MA, 1961).

\bibitem{laughlin} Laughlin S. B., {\it Z. Naturf.} {\bf 36} 910-912 (1981).

\bibitem{hateren} van Hateren~J.H.
{\it J. Comp. Physiology A} {\bf 171} 157-170, 1992.

\bibitem{atick_rev} Atick~J. J.{\it Network} {\bf 3} 213-251, 1992.

\bibitem{OF_nature} Olshausen B.A. \& Field D. J., 
{\it Nature} {\bf 381}, 607-609 (1996).

\bibitem{Baddeley} Baddeley R., {\it Cognitive Science}, in press (1997).


\bibitem{Carlson}  Carlson C. R., {\it Photog. Sci. Engng.} {\bf 22}  
69-71 (1978).

\bibitem{Burton} Burton G. J. and Moorhead I. R.,
{\it Appl. Opt.} {\bf 26} 157-170 (1987).  

\bibitem{Tollhurst} Tollhurst D. J., Tadmor Y. \& Tang Chao, 
{\it Ophthal. Physiol. Opt.} {\bf 12} 229-232 (1992).


\bibitem{BellSej-edges} Bell \& Sejnowski, 
{\it Vision Research} {\bf 37}, 3327-3338 (1997).

\bibitem{Field_goal} Field D. 
{\it Neural Computation} {\bf 6}, 559-601 (1994)


\bibitem{Benzi-multiaffine} Benzi, Biferale, Crisanti, Paladin, Vergassola 
\& Vulpiani,{\it Physica D} {\bf 65}, 352-358 (1993).



}


\end{thebibliography}
\end{document}